\documentclass[twocolumn,preprint]{elsart3p}

\usepackage{epsfig}
\usepackage{amssymb}
\usepackage{color,soul}

\begin{document}

\begin{frontmatter}



\title{A fully microscopic model of total level density in spherical nuclei}

\author[label1,label2]{N. Quang Hung \corauthref{cor1}},
\corauth[cor1]{Corresponding author.}
\ead{nguyenquanghung5@duytan.edu.vn}
\author[label3]{N. Dinh Dang},
\ead{dang@riken.jp}
\author[label1,label2]{L. Tan Phuc},
\ead{letanphuc191190@gmail.com}
\author[label4]{N. Ngoc Anh},
\ead{ngocanh8999@gmail.com}
\author[label1,label2]{T. Dong Xuan},
\ead{trandongxuan@gmail.com}
\author[label5]{T. V. Nhan Hao}
\ead{tvnhao@hueuni.edu.vn}

\address[label1]{Institute of Fundamental and Applied Sciences, Duy Tan University, Ho Chi Minh City 700000, Viet Nam}
\address[label2]{Faculty of Natural Sciences, Duy Tan University, Da Nang City 550000, Viet Nam}
\address[label3]{Quantum Hadron Physics Laboratory, RIKEN Nishina Center for Accelerator-Based Science, 2-1 Hirosawa, Wako City, 351-0198 Saitama, Japan}
\address[label4]{Dalat Nuclear Research Institute, Vietnam Atomic Energy Institute, 01 Nguyen Tu Luc, Dalat City 670000, Viet Nam}
\address[label5]{Faculty of Physics, University of Education, Hue University, 34 Le Loi Street, Hue City 530000, Viet Nam}

\begin{abstract}
A fully microscopic model for the description of nuclear level density (NLD) in spherical nuclei is proposed. The model is derived by combining the partition function of the exact pairing solution plus the independent-particle model at finite temperature (EP+IPM) with that obtained by using the collective vibrational states calculated from the self-consistent Hartree-Fock mean field with MSk3 interaction plus the exact pairing and random-phases approximation (SC-HFEPRPA). Two important factors are taken into account in a fully microscopic way, namely the spin cut-off and vibrational enhancement factors are, respectively, calculated using the statistical thermodynamics and partition function of the SC-HFEPRPA without any fitting parameters. The numerical test for two spherical $^{60}$Ni and $^{90}$Zr nuclei shows that the collective vibrational enhancement is mostly dominated by the quadrupole and octupole excitations. This is the first microscopic model confirming such an effect, which was phenomenologically predicted long time ago and widely employed in several NLD models. In addition, the influence of collective vibrational enhancement on nuclear thermodynamic quantities such as excitation energy, specific heat capacity and entropy is also studied by using the proposed model.
\end{abstract}

\end{frontmatter}

The concept of NLD, defined as the number of levels per unit of excitation energy, was first introduced a long time ago by Hans Bethe \cite{Bethe} by notifying that the number of excited states in atomic nuclei increases rapidly with the excitation energy. Consequently, it is impossible to individually treat those states, even by using advanced experimental and theoretical techniques. Thus, the NLD reflects the average properties of excited nuclei and has various applications in the study of not only nuclear structure and reactions but also nuclear engineering and astrophysics \cite{Rauscher}. The NLD also contains various information on the internal structure of atomic nuclei such as single-particle levels, pairing correlations, spin distributions, collective (vibrational and/or rotational) excitations, nuclear thermodynamics, etc. \cite{Ericson60}. Many theoretical studies have been carried out during the last seven decades in order to find a reliable and fully microscopic model of NLD.

In general, theoretical approaches to NLD are classified into the phenomenological and microscopic models. Phenomenological NLD models such as the back-shifted Fermi gas (BSFG) and constant temperature \cite{GC65,Dilg73} were derived based on simple analytical formulas containing some phenomenological parameters such as the level density parameter, shell correction, pairing energy, temperature, energy shift, spin cut-off factor, etc. The values of these parameters are obtained from the local and/or global fittings to a limited number of experimental data such as the experimental cumulative number of discrete levels at low-excitation energy $E^*$ and the neutron resonance data at $E^* = B_n$ with $B_n$ being the neutron binding energy \cite{Egidy05}. For nuclei, whose experimental NLD data are completely unknown, the prediction of the above models becomes questionable (see e.g., Ref. \cite{Voinov07}). In this case, the development of microscopic methods should be more favorable. Several microscopic NLD models have been developed such as the Hartree-Fock BCS (HFBCS) \cite{HFBCS}, static path approximation (SPA) \cite{SPA}, SPA plus random-phase approximation (SPA+RPA) \cite{SPARPA,SPARPA1}, finite-temperature shell model Monte Carlo (SMMC) \cite{SMMC,SMMC1}, and Hartree-Fock-Bogolyubov plus combinatorial method (HFBC) \cite{HFBC,HFBC1}. They were often derived based on a Hamiltonian, which consists of a realistic or phenomenological single-particle/mean-field potential of Woods-Saxon (SPA+RPA or SMMC) or Skyrme Hartree-Fock (HFBCS) or Hartree-Fock-Bogolyubov (HFBC) combined with the residual interactions (pairing correlation and collective excitations) beyond the single-particle mean field. However, some ambiguities within the above microscopic models still remain. Firstly, pairing is approximately treated within the HFBCS and HFBC, which both violate the particle-number conservation. Consequently, the NLD obtained within the HFBCS and HFBC has to be normalized by using two additional parameters, whose values are extracted by fitting to the experimental cumulative number of discrete levels at low $E^*$ and neutron resonance data at $E^* = B_n$ (see e.g., Eq. (9) of Ref. \cite{HFBC1}), losing certainly their microscopic nature. Secondly, the spin cut-off factor $\sigma$, which is important for determining the spin distribution of the NLD, is empirically determined from the rigid-body limit within the HFBCS, HFBC, SPA, and SPA+RPA, whereas it is microscopically calculated based on the exact ratio between the total $\rho(E^*)$ and spin projected $\rho_J(E^*)$ NLDs within the SMMC. Thirdly, the residual correlations are either not taken into account (within the HFBCS and HFBC) or taken into account in a simplified way (within the SPA+RPA). For instance, the HFBC considers only excitations built on all uncorrelated particle-hole configurations, whereas the SPA+RPA employs very simple two-body interactions of the quadrupole-quadrupole and/or higher excitations, (see e.g., Eq. (1) of Ref. \cite{SPARPA1}). Lastly, the numerical calculations within the SPA+RPA and SMMC are time consuming, in particular for heavy nuclei.

Recently, we have proposed a microscopic NLD method based on the exact pairing plus independent-particle model at finite temperature (EP+IPM) \cite{EPIPM}. This model, which contains no fitting parameters to the experimental NLD data and has very short computing time, has provided a good description for the NLDs and thermodynamic properties of not only hot $^{170-172}$Yb \cite{EPIPM} and $^{60-62}$Ni \cite{EPIPM1} nuclei but also hot rotating $^{96}$Tc \cite{EPIPM2}, $^{184}$Re, $^{200}$Tl, $^{211}$Po, and $^{212}$At \cite{EPIPM3} isotopes. However, it still contains two shortcomings. The first shortcoming is the spin cut-off factor taken from the empirical formula in the rigid-body limit for axially deformed nuclei, namely $\sigma_\perp \approx 0.015A^{5/3}T$ and $\sigma_\parallel = \sigma_\perp\sqrt{(3-2\beta_2)/(3+\beta_2)}$, where $T, \beta_2$, $\sigma_\perp$, and $\sigma_\parallel$ are nuclear temperature, quadrupole deformation parameter, perpendicular and parallel spin cut-off factors, respectively \cite{EPIPM}. The second shortcoming is the collective enhancement factor $k_{\rm coll}$ consisting the vibrational $k_{\rm vib}$ and rotational $k_{\rm rot}$ excitations, which are also empirically described as 
\begin{eqnarray}
&&k_{\rm vib} = \exp[0.055A^{2/3}T^{4/3}] ~, \label{kvibemp} \\ 
&& k_{\rm rot} = (\sigma_\perp^2 - 1)/[1+e^{(E^*-U_C)/D_C}]+1 \label{krot} ~,
\end{eqnarray} 
where $D_C = 1400\beta_2^2A^{-2/3}$ and $U_C = 120\beta_2^2A^{1/3}$. The second shortcoming is the most difficult problem of the present NLD models. Regarding the spin cut-off factor, it can be microscopically calculated by using the statistical thermodynamic method e.g., in Refs. \cite{Moretto72,Bekhami73}, namely, 
\begin{equation}
\sigma^2 = \frac{1}{2}\sum_k m_k^2{\rm sech}^2\frac{1}{2}\beta E_k ~, \label{sigma}
\end{equation}
where $m_k$ is the single-particle spin projection (written in the deformed basis), $\beta = 1/T$ is the inverse of temperature $T$, and the quasiparticle energy $E_k$ is calculated from $E_k = \sqrt{(\epsilon_k - \lambda)^2 + \Delta^2}$ if pairing is included or $E_k = \epsilon_k - \lambda$ if no pairing is considered ($\Delta$, $\epsilon_k$, and $\lambda$ are pairing gap, single-particle energies and chemical potential, respectively). Within the EP, the pairing gap $\Delta$ and quasiparticle energy $E_k$ are exactly calculated by using e.g., Eqs. (11) and (12) of Ref. \cite{DangPRC12}. As for the vibrational enhancement, there exists another approximate formula given in Refs. \cite{HFBC,Grudzevich88,Techdoc06}
\begin{equation}
k_{\rm vib} = \exp[\delta S - \delta U/T)~, \label{kvib}
\end{equation}
where $\delta S = \sum_i (2\lambda_i + 1)[(1+n_i){\rm ln}(1+n_i)-n_i{\rm ln}n_i]$ and $\delta U = \sum_i (2\lambda_i +1)\omega_i n_i$ are the changes in the entropy and excitation energy, respectively, due to vibrational modes with $\omega_i$, $\lambda_i$, and $n_i$ being the energies, multipolarities, and temperature-dependent occupation numbers, respectively. The occupation numbers $n_i$ are defined as $n_i = \exp(-\gamma_i/2\omega_i)/[\exp(\omega_i/T)-1]$ with $\gamma_i = 0.0075A^{1/3}(\omega_i^2+4\pi^2T^2)$ being the spreading widths of the vibrational excitations. As for the phonon energies $\omega_i$, the modified equations including shell correction $E_{\rm shell}$ are considered, namely $\omega_2 = 65A^{-5/6}/(1+0.05E_{\rm shell})$ for the quadrupole and $\omega_3 = 100A^{-5/6}/(1+0.05E_{\rm shell})$ for the octupole \cite{HFBC}. It is clear to see that Eq. (\ref{kvib}) adopts only the lowest energies $\omega_{2,3}$ of the quadrupole ($\lambda = 2$) and octupole ($\lambda = 3$) vibrations and neglects higher vibrational energies and other multipolarities such as monopole ($\lambda = 0$), dipole ($\lambda = 1$), hexadecupole ($\lambda = 4$), etc. Indeed, such vibrational energies and multipolarities can be microscopically calculated within the random-phase approximation (RPA), one of the most extensive approximations for nuclear collective vibrational excitations. In this case, one should construct a vibrational partition function of the following form
\begin{equation}
Z_{\rm vib}(T) = \sum_\lambda (2\lambda + 1){\sum_i}e^{-E_i^\lambda/T} ~, \label{Zvib}
\end{equation}
in the canonical ensemble, where $E_i^\lambda$ are all the eigenvalues (energies) obtained by solving the RPA equation for the corresponding multipolarity $\lambda$, which runs from 0 to 4 or 5. Similarly, one can construct a partition function for rotational excitation $Z_{\rm rot}(T)$ similar to Eq. (\ref{Zvib}) by replacing the energies $E_i^\lambda$ with rotational states obtained from different rotational bands. 

In the present Letter, we develop a fully microscopic method for the description of NLD, limited to spherical nuclei (no rotational enhancement or $k_{\rm rot} = 1$), although the proposed idea is applicable also for deformed systems. The model is derived based on the EP+IPM in Ref. \cite{EPIPM}, however, three significant improvements have been proposed. First, the single-particle spectra are taken from the Hartree-Fock mean field plus exact pairing (HF+EP) with an effective Skyrme interaction (MSk3) as developed in Ref. \cite{HFEP}, instead of the phenomenological Woods-Saxon potential. This HF+EP with MSk3 force has provided a very good description of not only binding and two nucleon-separation energies but also nucleon densities and single-particle occupation numbers of light and spherical $^{22}$O and $^{34}$Si nuclei at zero and finite temperatures. Second, the spin cut-off parameter is calculated using Eq. (\ref{sigma}), instead of empirical formula. Third, the vibrational enhancement is microscopically treated by combining the EP+IPM partition function with that given in Eq. (\ref{Zvib}), in which the self-consistent Hartree-Fock+EP+RPA (SC-HFEPRPA) with the same MSk3 force \cite{EPRPA} is used instead of the conventional HF+RPA, namely 
\begin{equation} 
\ln Z'_{\rm total}(T) = \ln Z'_{\rm EP+IPM}(T) + \ln Z'_{\rm SC-HFEPRPA}(T) ~, \label{Ztol}
\end{equation}
where $Z'(T)$ denotes the excitation partition function \cite{IPM}. Knowing the total partition function (\ref{Ztol}), one can easily calculate the excitation energy $E(T)$, entropy $S(T)$, and heat capacity $C(T)$, which are later used together with the spin cut-off factor $\sigma$ to calculate the total NLD $\rho_{\rm tot}(E^*)$ as has been done in Ref. \cite{EPIPM}. Thus, the collective vibrational enhancement is directly included in the total partition function and the vibrational enhancement factor $k_{\rm vib}$ can be calculated via $k_{\rm vib}(E^*) = \rho_{\rm tot}(E^*)/\rho_{\rm int}(E^*)$, where $\rho_{\rm int}(E^*)$ is the intrinsic NLD, that is, the NLD obtained by using $Z_{\rm EP+IPM}$ only. The model in this case is fully microscopic as it does not contain any empirical expressions and fitting parameters to the NLD data. 

\begin{figure}
       \includegraphics[scale=0.4]{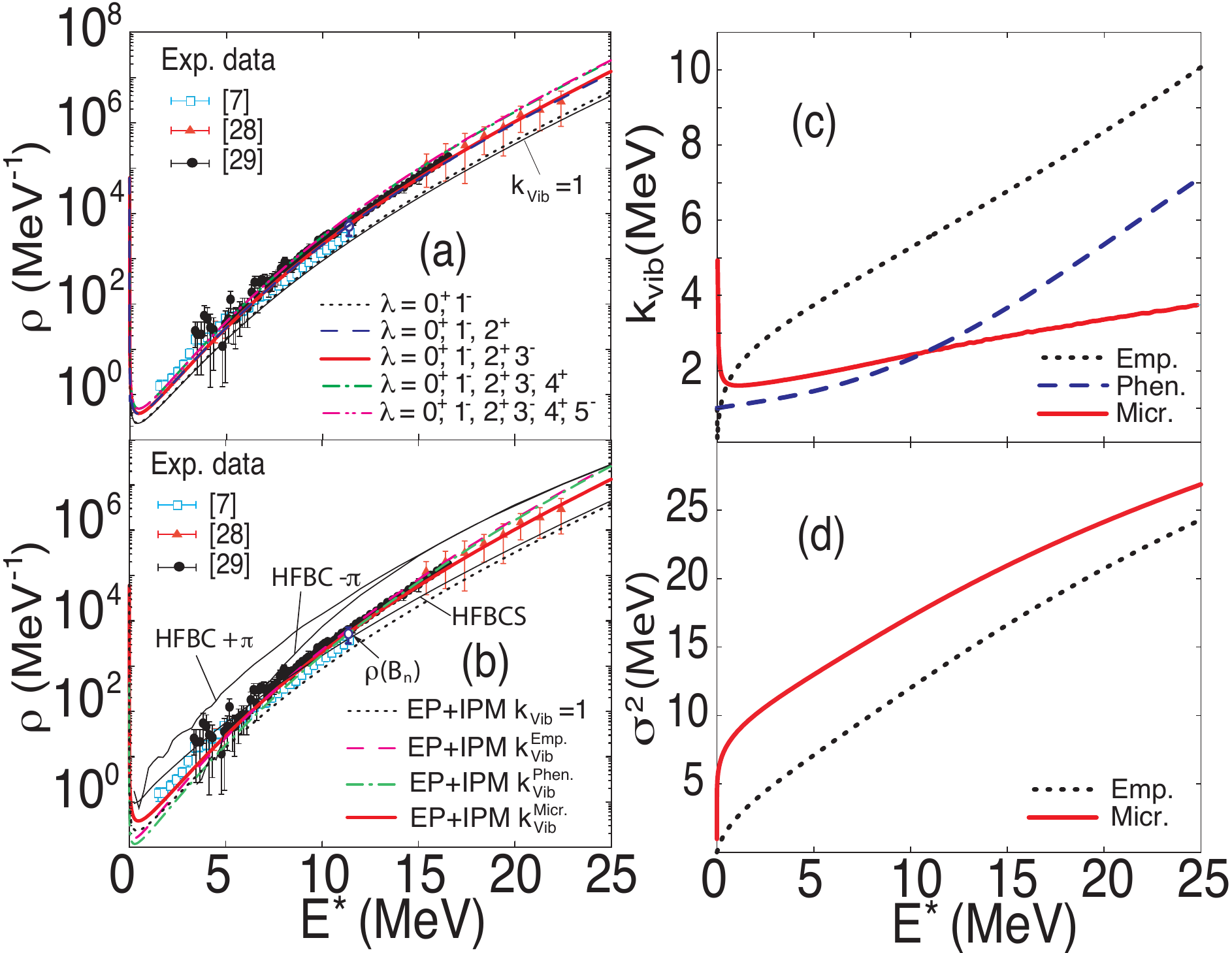}
       \caption{(Color online) (a) The total NLD obtained within the fully microscopic EP+IPM by gradually adding higher collective vibrational modes $\lambda$ to the vibrational partition function (\ref{Zvib}) versus the experimental data for $^{60}$Ni. (b) The best total NLD obtained from (a) with $\lambda = 0^+, 1^-, 2^+$, and $3^-$ (thick solid line) in comparison with those obtained within the HFBCs (for positive $+\pi$ and negative $-\pi$ parities and HFBCS as well as those obtained within the EP+IPM with phenomenological ($k_{\rm vib}^{\rm Phen.}$), empirical ($k_{\rm vib}^{\rm Emp.}$), and without ($k_{\rm vib} = 1$) vibrational enhancements. (c) The vibrational enhancement factor ($k_{\rm vib}$) obtained by using empirical (\ref{kvibemp}), phenomenological (\ref{kvib}), and microscopic formulas. (d) The spin cut-off parameter $\sigma^2$ obtained by using empirical (rigid-body limit) and microscopic (\ref{sigma}) formulas.
        \label{fig1}}
    \end{figure}

\begin{figure}
       \includegraphics[scale=0.4]{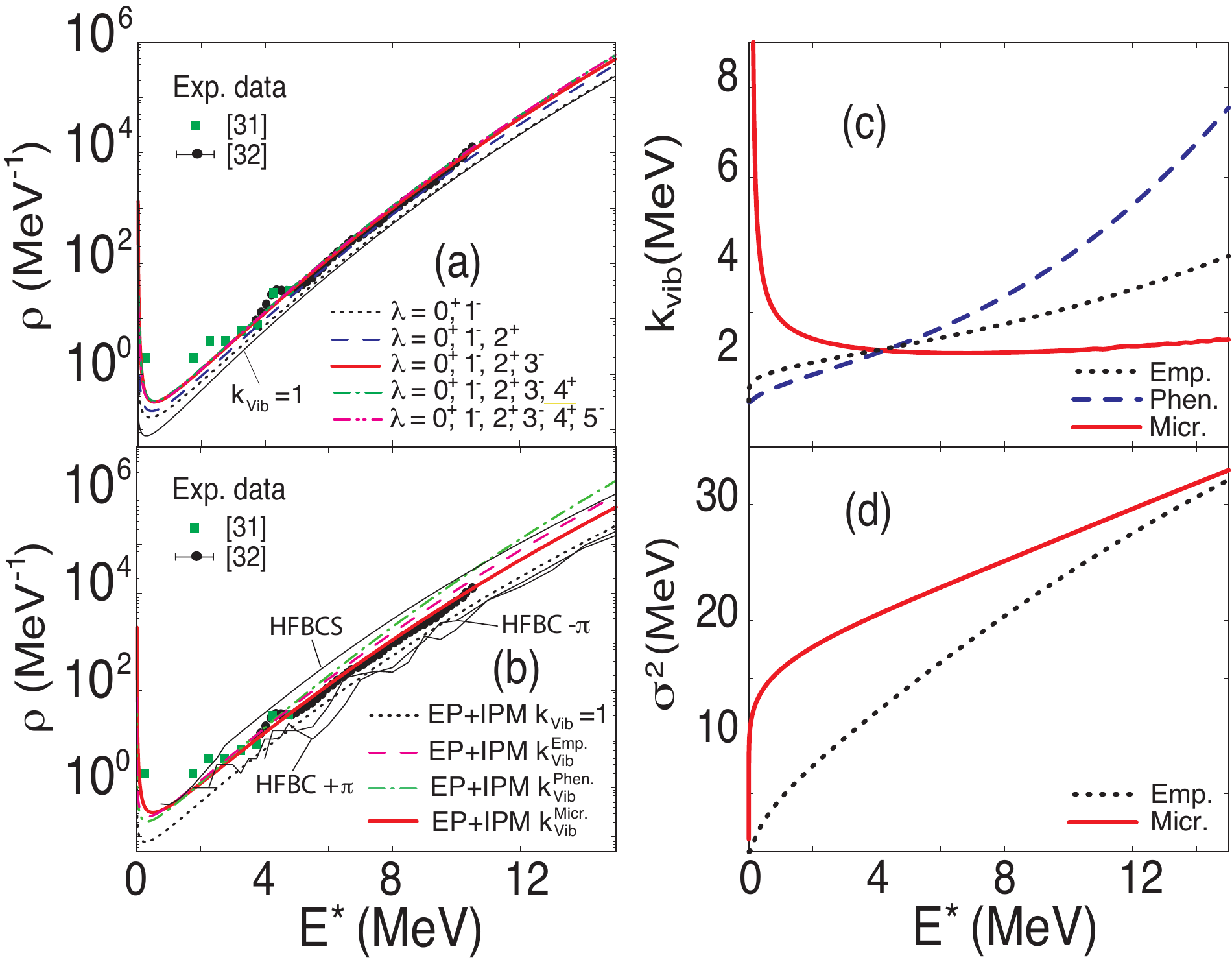}
       \caption{(Color online) Same as Fig. \ref{fig1} but for $^{90}$Zr.
        \label{fig2}}
    \end{figure}

\begin{figure}
       \includegraphics[scale=0.4]{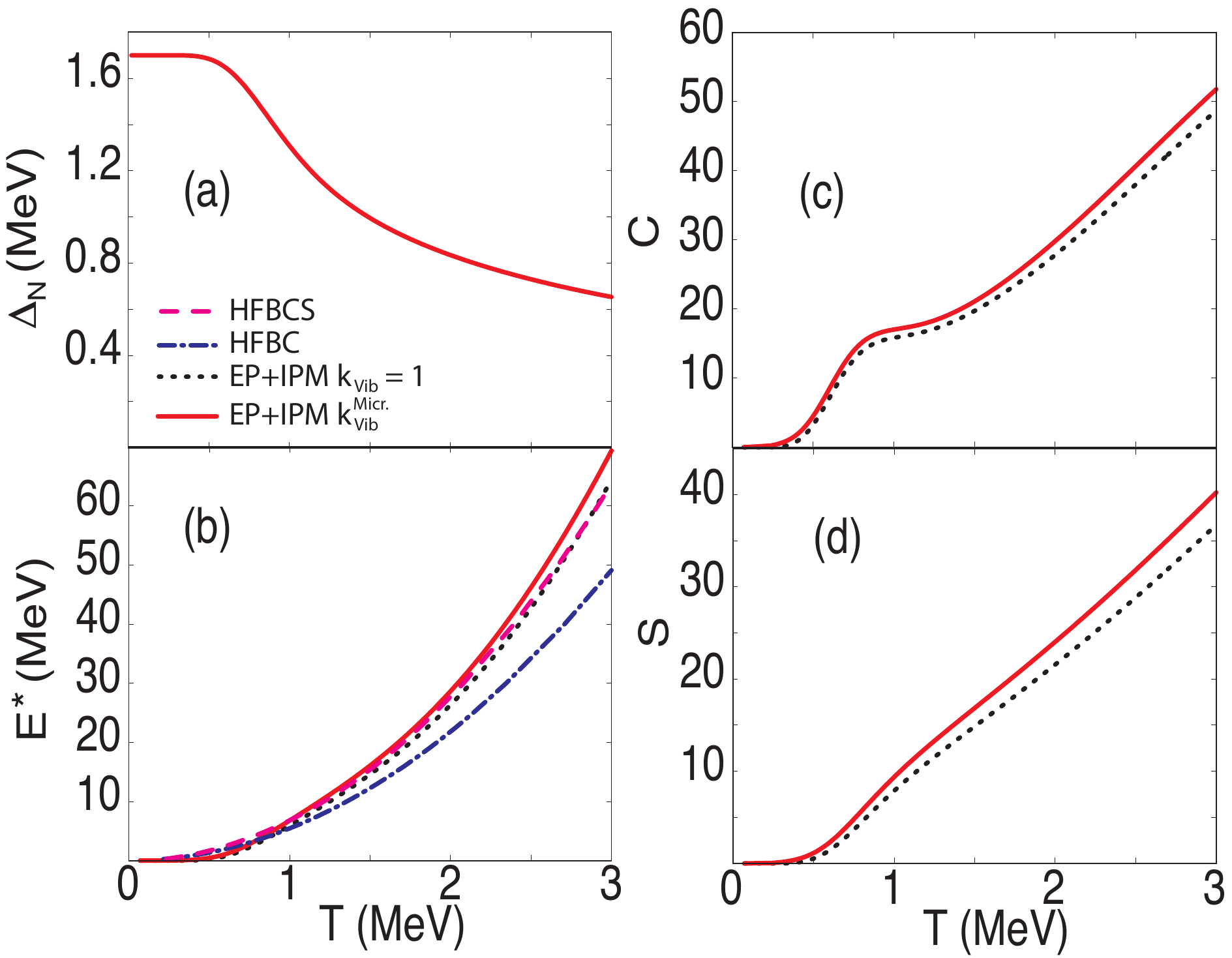}
       \caption{(Color online) The neutron pairing gap $\Delta_N$ (a), excitation energy $E^*$ (b), heat capacity $C$ (c), and entropy $S$ (d) as functions of $T$ obtained within the EP+IPM for $^{60}$Ni with (solid line) and without (dotted line) collective vibrational enhancement. In (b), the excitation energy $E^*$ obtained within the HFBC and HFBCS are also plotted, for comparison with the EP+IPM calculations (with and without collective enhancement).
        \label{fig3}}
    \end{figure}

To test the proposed model, we select the spherical $^{60}$Ni nucleus, which is the only nucleus whose experimental NLD data are presently available from $E^* \sim 0$ up to $\sim 23$ MeV, well above $E^* = B_n = 11.358$ MeV \cite{Voinov07,exp1,exp3}. The high-energy part of the NLD is very important to test the validity of the microscopic calculations. The average binding energy BE/A and energy $E_{2_1^+}$ of the first $2^+$ state obtained within the SC-HFEPRPA calculation for $^{60}$Ni are -8.736 and 1.336 MeV, respectively, which are in excellent agreement with the experimental data (BE/A = -8.780 MeV and $E_{2_1^+}$ = 1.333 MeV). Moreover, the energy weighted sum rules for the isoscalar (IS) and isovector (IV) excitations are perfectly conserved for all the multipolarities $\lambda^\pi = 0^+, 1^-, 2^+, 3^-, 4^+$, and $5^-$ with natural parities. The results obtained are shown in Figs. \ref{fig1} and \ref{fig3}. A similar test for another spherical $^{90}$Zr nucleus is also performed and the results are illustrated in Fig. \ref{fig2}. The pairing interaction $G$ is selected, as usual, so that the neutron and proton pairing gaps at $T=0$ agree with those obtained from the experimental odd-even mass differences for $^{60}$Ni and $^{90}$Zr, respectively \cite{EPIPM}.


Fig. \ref{fig1}(a) depicts the total NLDs $\rho(E^*)$ obtained within our fully microscopic EP+IPM by gradually adding the higher multipolarities $\lambda$ to the vibrational partition function (\ref{Zvib}). It is clear to see in this figure that the monopole and dipole excitations ($\lambda = 0^+$ and $1^-$) (dotted line) have very small enhancement  to $\rho_{\rm int}$ (thin solid line). Adding the quadrupole $\lambda = 2^+$ vibration significantly increases the NLD and the result obtained (dash line) agrees well with not only the low-energy ($E^* < B_n$) but also the high-energy ($E^* \geq B_n$) NLD data. Adding the octupole $\lambda = 3^-$ excitation (thick solid line) does not enhance much the NLD. However, adding higher hexadecapole $\lambda = 4^+$ (dash dotted line) and tetrahedral $\lambda = 5^-$ (dash-dot dotted line) excitations enhances the NLD further but the results obtained slightly overestimates the experimental data, in particular the data at $E^* \geq B_n$. This is because their excitations always take place at a very high excitation energy, which goes beyond the vibrational excitation region and thus, should not be included in the vibrational partition function. The analysis in Fig. \ref{fig1}(a) strongly indicates that the vibrational enhancement of NLD is mainly due to the quadrupole and octupole excitations. Indeed, this suggestion was initiated long time ago in Ref. \cite{Grudzevich88} but without any microscopic justification. Since then, it was widely used in various NLD models, e.g., Refs. \cite{HFBC, Techdoc06}. Therefore, the results shown in Fig. \ref{fig1}(a) are of particular valuable as they are the first microscopic calculation, which confirms the important role of the quadrupole and octupole excitations in the NLD.

In Fig. \ref{fig1}(b), the best NLD obtained within the EP+IPM with $\lambda = 0^+, 1^-, 2^+$, and $3^-$ is compared with those obtained within other microscopic HFBCS and HFBC (for positive $+\pi$ and negative -$\pi$ parities) approaches taken, respectively, from RIPL-2 \cite{RIPL2} and RIPL-3 \cite{RIPL3} nuclear database as well as those calculated within the EP+IPM without vibrational enhancement ($k_{\rm vib} = 1$) and with empirical (\ref{kvibemp}) and phenomenological (\ref{kvib}) formulas for $k_{\rm vib}$. It is seen that the HFBC over estimates the data, whereas the HFBCS agrees only with the data in Ref. \cite{Voinov07} (open squares) at $E^* >$ $\sim$4 MeV because this method was normalized to fit to these data. Once the data are updated and extended to higher $E^*$ as in Ref. \cite{exp3}, the HFBCS NLD deviates from the new data. This result clearly shows the main drawback of the HFBC and HFBCS models. The EP+IPM NLDs, which employ the empirical $k_{\rm vib}^{\rm Emp.}$ and phenomenological $k_{\rm vib}^{\rm Phen.}$, agree with the data at 5 MeV $< E^* < B_n$ only due to the fact that the values of $k_{\rm vib}^{\rm Emp.}$ and $k_{\rm vib}^{\rm Phen.}$ are larger than the corresponding microscopic calculation $k_{\rm vib}^{\rm Micr.}$ [Fig. \ref{fig1}(c)]. The later is calculated by taking the ratio between the EP+IPM NLD with $\lambda = 0^+, 1^-, 2^+$, and $3^-$ and its state density ($k_{\rm vib} = 1$).\footnote{It is seen in Fig. \ref{fig1}(c) that the value of $k_{\rm vib}^{\rm Micr.}$ does not start from 1 at $E^* \rightarrow 0$ as those of $k_{\rm vib}^{\rm Emp.}$ and $k_{\rm vib}^{\rm Phen.}$. This is due to the well-known unphysical divergence of the saddle-point approximation at very low $T$ or $E^*$, which has been widely employed in most microscopic NLD models \cite{BM}.} Moreover, the microscopic spin cut-off factor calculated within the EP+IPM using Eq. (\ref{sigma}) is found to be larger than that calculated from the empirical formula of rigid body [Fig. \ref{fig1}(d)]. Similar results can be seen in Fig. \ref{fig2} but for the $^{90}$Zr nucleus, whose NLD data are available below $E^* < B_n$ \cite{Byun14}. The EP+IPM NLD with the quadrupole and octupole excitations is found in the best agreement with the experimental data [Fig. \ref{fig2}(a)], whereas the HFBCS (HFBC) overestimates (underestimates) the data [Fig. \ref{fig2}(b)]. The values of $k_{\rm vib}^{\rm Emp.}$ and $k_{\rm vib}^{\rm Phen.}$ are higher than that of $k_{\rm vib}^{\rm Micro.}$ [Fig. \ref{fig2}(c)], while the microscopic spin cut-off factor is higher than the empirical one [Fig. \ref{fig2}(d)], similar to those obtained for $^{60}$Ni in Fig. \ref{fig1}. All the results shown in Figs. \ref{fig1} and \ref{fig2} are particularly important because it ensures the validity of our fully microscopic NLD model, in which pairing is exactly treated and the spin cut-off and collective vibrational enhancement factors are microscopically calculated.

Because of the collective excitations, not only the NLD but also the thermodynamic quantities of excited nuclei are enhanced. We show in Fig. \ref{fig3} the thermodynamic quantities such as neutron pairing gap $\Delta_N$ (a), excitation energy $E^*$ (b), heat capacity $C$ (c), and entropy $S$ (d) as functions of $T$ obtained within the EP+IPM with and without the contribution of collective vibrational enhancement. For $E^*(T)$, the predictions of the HFBC and HFBCS taken from Refs. \cite{RIPL2,RIPL3} are also plotted in Fig. \ref{fig3}(b). Obviously, the neutron pairing gap $\Delta_N$ (proton pairing gap is zero because $^{60}$Ni has a proton magic number) calculated within the exact pairing decreases with increasing $T$ and remains finite even at $T = 3$ MeV, in agreement with many microscopic calculations (see e.g., Ref. \cite{HungROPP19}). Consequently, one can see an $S$-shaped heat capacity, which indicates the signature of superfluid-normal phase transition in finite nuclear systems. The collective vibrational enhancement is seen to enhance all the thermodynamic quantities, except the exact pairing gap, which is the concept of the mean field, whereas the collective enhancement goes beyond the mean-field concept. It is also seen that the HFBC predicts a rather small excitation energy in comparison with the HFBCS and our EP+IPM [Fig. \ref{fig3}(b)]. This result of HFBC can be easily understood because this method is constructed based only on all the combinations of uncorrelated particle-hole excitations, meaning that some correlated excitations are not taken into account \cite{HFBC,HFBC1}. The HFBCS excitation energy is very close to that of the EP+IPM calculated without collective enhancement. This result is reasonable because the HFBCS does not explicitly treat the collective enhancement \cite{HFBCS}. The results shown in Fig. \ref{fig3} are very interesting because they are obtained from the first microscopic model exploring the effect of collective enhancement on nuclear thermodynamic quantities.

The present Letter proposes a fully microscopic model for the description of total NLD. The model is proposed by combining the thermodynamic partition function of the exact pairing solution with that obtained within the finite-temperature independent-particle model (EP+IPM) and collective vibrational excitation modes. The later are calculated from the Hartree-Fock mean field with MSk3 interaction and self-consistently combined with the exact pairing solution and random-phase approximation (SC-HFEPRPA). In addition, the spin cut-off parameter is also microscopically calculated within the EP+IPM using the statistical thermodynamics. The numerical test has been carried out for the spherical $^{60}$Ni nucleus, the only nucleus whose NLD data are available from the excitation energy of 0 to about 23 MeV. A similar test has also been performed for the spherical $^{90}$Zr nucleus, whose NLD data are provided below the neutron binding energy. The results obtained show that by combining the EP+IPM partition function with that obtained using the collective vibrational states taken from the SC-HFEPRPA calculation, which excellently reproduces the experimental binding energy and energy of the first $2^+$ state, we are able to study the contributions of different vibrational modes from the monopole ($0^+$) to tetrahedral ($5^-$) states, from which the quadrupole and octupole excitations are found to be the most importance. This is, indeed, the first microscopic model capable to confirm the important role of the vibrational quadrupole and octupole excitations in the NLD. The NLD obtained within this model is found in a much better agreement with the experimental data than those calculated within other approaches such as HFBCS and HFBC as well as those obtained by using the empirically/phenomenologically vibrational enhancement and spin cut-off factors. It has also been found that the vibrational enhancement factor obtained within our fully microscopic approach is lower than that calculated using the empirical and phenomenological formulas. Regarding the spin cut-off factor, its value obtained within our microscopic model is larger than that obtained by using the empirical formula. The effect of vibrational enhancement on nuclear thermodynamic quantities such as excitation energy, entropy, and heat capacity is also studied, for the first time, within our fully microscopic model. Finally, the present model does not consume much computing time, namely the EP+IPM calculation takes less than 5 min \cite{EPIPM}, whereas the SC-HFEPRPA takes about 1 hour for the calculation of each multipolarity. Therefore, one calculation for one nucleus, even a heavy one, takes less than 6 hours and thus can be performed on a PC. Although the present Letter is restricted to the description of NLD in spherical nuclei, the proposed method can also be applied to deformed isotopes, that is, the rotational enhancement factor can also be microscopically calculated from the partition function of all the excited states coming from the rotational bands. The result obtained for deformed nuclei will be reported in the forthcoming papers. 

\section*{Acknowledgement}
This work is funded by the National Foundation for Science and Technology Development (NAFOSTED) of Vietnam through Grant No. 103.04-2019.371.

\end{document}